\documentclass{article}
\usepackage{icmc2025_paper_template}
\usepackage{times}
\usepackage{ifpdf}
\usepackage{soul}
\usepackage[english]{babel}
\usepackage{amsmath}
\usepackage{subfig}

\def\papertitle{\text{ReMi}: A Random Recurrent Neural Network Approach to Music Production}
\def\firstauthor{Hugo Chateau-Laurent}
\def\secondauthor{Tara Vanhatalo}
\def\thirdauthor{Wei-Tung Pan}
\def\fourthauthor{Xavier Hinaut}

\newif\ifpdf
\ifx\pdfoutput\relax
\else
   \ifcase\pdfoutput
      \pdffalse
   \else
      \pdftrue
  \fi
\fi

\ifpdf 
  \usepackage[pdftex,
    pdftitle={\papertitle},
    pdfauthor={\firstauthor, \secondauthor, \thirdauthor, \fourthauthor},
    bookmarksnumbered, 
    pdfstartview=XYZ 
   ]{hyperref}

  \usepackage[pdftex]{graphicx}
  \graphicspath{{./figures/}}
  \DeclareGraphicsExtensions{.pdf,.jpeg,.png}

  \usepackage[figure,table]{hypcap}

\else 
  \usepackage[dvips,
    bookmarksnumbered, 
    pdfstartview=XYZ 
  ]{hyperref}  

  \usepackage[dvips]{epsfig,graphicx}
  \graphicspath{{./figures/}}
  \DeclareGraphicsExtensions{.eps}

  \usepackage[figure,table]{hypcap}
\fi

\hypersetup{
    colorlinks,%
    citecolor=black,%
    filecolor=black,%
    linkcolor=black,%
    urlcolor=black
}

\title{\papertitle}

\fourauthors
  {1in}
  {\firstauthor} {Allendia, Inria centre of Bordeaux University \\ %
    {\tt \href{mailto:hugo.chateau-laurent@allendia.com}{hugo.chateau-laurent@allendia.com}}}
  {\secondauthor} {Allendia, Inria centre of Bordeaux University}
  {\thirdauthor} {Inria centre of Bordeaux University}
  {\fourthauthor} {Inria centre of Bordeaux University, IMN, LaBRI}

\begin{document}
\capstartfalse
\maketitle
\capstarttrue
\begin{abstract}
Generative artificial intelligence raises concerns related to energy consumption, copyright infringement and creative atrophy. We show that randomly initialized recurrent neural networks can produce arpeggios and low-frequency oscillations that are rich and configurable. In contrast to end-to-end music generation that aims to replace musicians, our approach expands their creativity while requiring no data and much less computational power. More information can be found at:
 \url{https://allendia.com/}

\end{abstract}

\section{Introduction}\label{sec:introduction}
Artificial intelligence continues to drive significant changes in music production. However, current methods often require vast amounts of high-quality data, which are not always readily available. These models also tend to copy the statistical patterns of existing compositions, raising concerns over copyright infringement and stifling of creativity \cite{doshi2024generative}. Moreover, the energy demands of training large networks pose significant environmental concerns, as these processes are computationally intensive.

We present ReMi (for Reservoir MIDI)
, a suite of music production tools designed to assist musicians without relying on extensive data or replacing human creativity. ReMi employs untrained, randomly initialized recurrent neural networks (RNNs) \cite{jaeger2004harnessing} to generate configurable signals. While ReMi does not include a learning phase in the traditional machine learning sense, we argue that its generative capacity stems from the intrinsic computational dynamics of Echo State Networks (ESNs). The current suite includes an arpeggiator and a Low-Frequency Oscillator (LFO), both allow users to explore a wide range of possibilities by tuning hyperparameters in real time.
By decoupling signal generation from data-driven training, ReMi produces outputs that are not tied to specific genres or styles, addressing concerns about the homogenization of AI-based creations \cite{doshi2024generative}. We argue that this approach mitigates risks to creativity and represents a more sustainable use of neural networks in music production.

\section{Echo State Networks}
ReMi builds upon ESNs \cite{jaeger2004harnessing}, an instance of reservoir computing. The state $\boldsymbol{h}$ of these networks is updated as follows:
\begin{align}
	\boldsymbol{h}_t &= \text{tanh} \bigg( \boldsymbol{W}^{in}\boldsymbol{x}_t + \boldsymbol{W}\boldsymbol{s}_{t-1} + \boldsymbol{W}^{fb}\boldsymbol{y}'_{t-1} + \boldsymbol{b} \bigg) \\
	\boldsymbol{s}_t &= (1-\alpha) \boldsymbol{s}_{t-1} + \alpha \boldsymbol{h}_t \\
	\boldsymbol{y}_t &= \boldsymbol{W}^{out}\boldsymbol{s}_t,
	\label{esn_update}
\end{align}
where $\boldsymbol{W}^{in}$, $\boldsymbol{W}$, $\boldsymbol{W}^{fb}$ and  $\boldsymbol{W}^{out}$ are the input, recurrent, feedback and output weights respectively,  $\boldsymbol{b}$ is the bias, $\boldsymbol{x}$ is the input and $0 \leq \alpha \leq 1$ is the leak rate. Typically, $\boldsymbol{y}'=\boldsymbol{y}$, but we introduce this distinction to allow for additional application-specific computation. We will also see that inputs are not necessarily needed. Weights are randomly initialized and kept fixed, and the user indirectly sculpts the signals by modifying network parameters. This replaces supervised learning with a human-in-the-loop, trial-and-error process, turning  the parameter space into a space of creative exploration.

\section{Low-Frequency Oscillator}\label{sec:LFO}

We first demonstrate that ESNs are well suited to implement parametric LFOs.
The most simple implementation involves no external input. 
The output is one-dimensional and can be mapped to modulate arbitrary external variables. In this case, we apply the sigmoid function to the output:
\begin{align}
    \boldsymbol{y}'_t = (1+e^{-\boldsymbol{y}_t})^{-1}
\end{align}
The user can control various hyperparameters in real time, e.g. the scale of the weight matrices and leak rate. 
If no desired dynamics are found, the user can choose to reset the network parameters and change the number of neurons. 

\begin{figure}
\centering
\subfloat[Sine]{
    \includegraphics[width=0.35\columnwidth, trim={0 1cm 0 1.8cm}, clip]{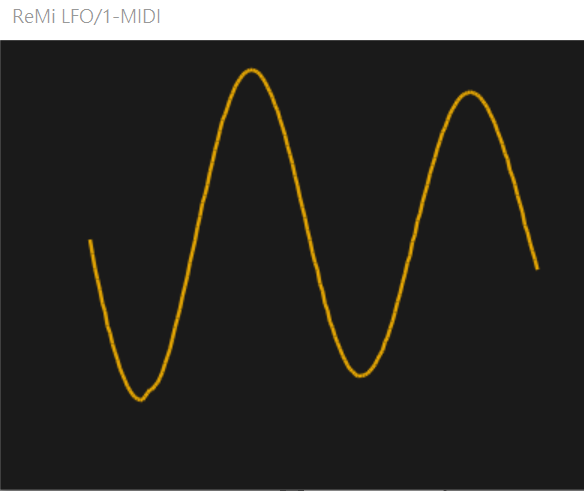}
    \label{fig:}
    }
\hspace{20pt}
\subfloat[Square]{
    \includegraphics[width=0.35\columnwidth, trim={0 1cm 0 1cm}, clip]{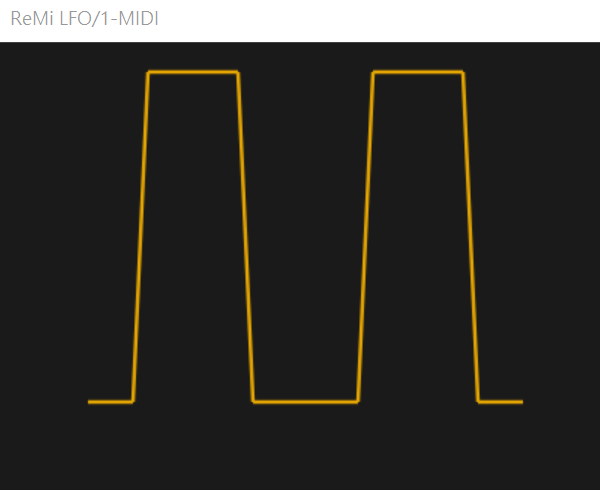}
    \label{fig:}
    }
    
\subfloat[Sawtooth]{
    \includegraphics[width=0.35\columnwidth, trim={0 1cm 0 1cm}, clip]{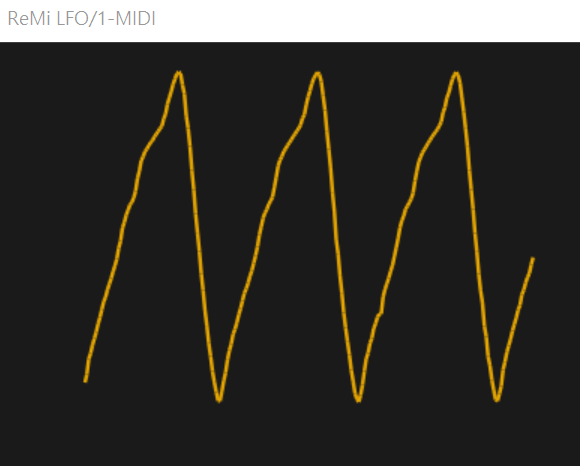}
    \label{fig:}
    }
\hspace{20pt}
\subfloat[Triangle]{
    \includegraphics[width=0.35\columnwidth, trim={0 1cm 0 1cm}, clip]{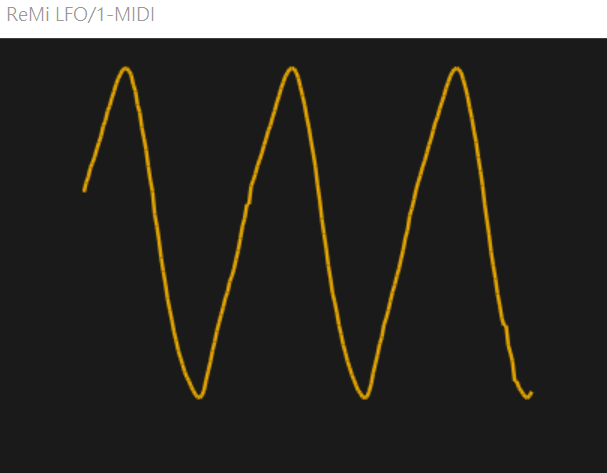}
    \label{fig:}
    }
    
\subfloat[Complex 1]{
    \includegraphics[width=0.35\columnwidth, trim={0 0.5cm 0 1cm}, clip]{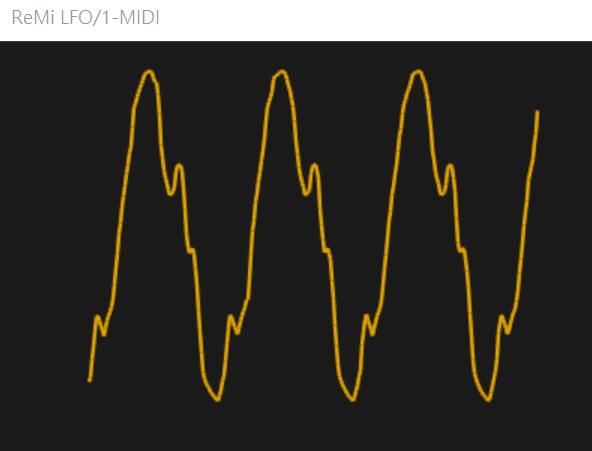}
    \label{fig:}
    }
\hspace{20pt}
\subfloat[Complex 2]{
    \includegraphics[width=0.35\columnwidth, trim={0 0.5cm 0 1cm}, clip]{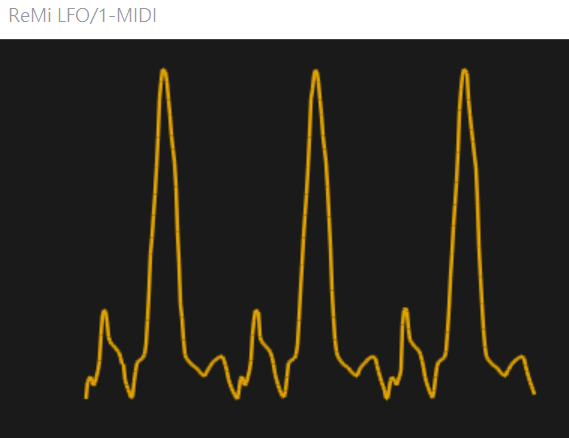}
    \label{fig:}
    }
\caption{\label{fig:waves}{\it \small LFO-generated common and complex waveforms.}}
\end{figure}

Example waveforms generated with ReMi are shown in Fig. \ref{fig:waves}. 
We reproduce common waveforms empirically to show that traditional oscillations can be created. 
Complex patterns can be found more easily than common waveforms. 
All example oscillations were not rigid and can be modified in a ``continuous'' manner by tuning the hyperparameters.

\section{Arpeggiator}
\label{sec:arp}

\begin{figure}
\subfloat[Principal component analysis]{
    \includegraphics[width=0.4\columnwidth, trim={0 0 0 0}, clip]{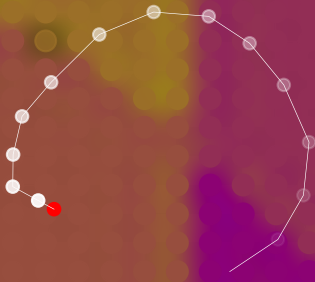}
    \label{fig:}
    }
\hfill
\subfloat[Neural activity]{
    \includegraphics[width=0.44\columnwidth, trim={0 0 0 0}, clip]{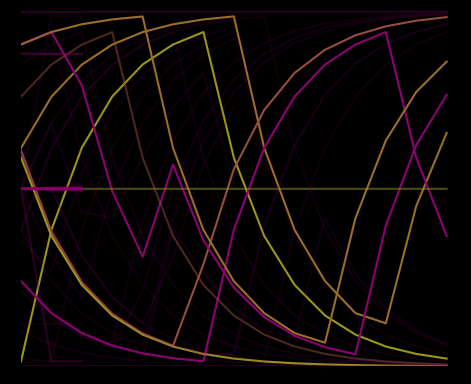}
    \label{fig:}
    }
\hfill
\centering
\subfloat[Network connectivity and activity. Vertices represent neurons scaled by activity. Edges represent weights.]{
    \includegraphics[width=0.5\columnwidth, trim={0 0 0 0}, clip]{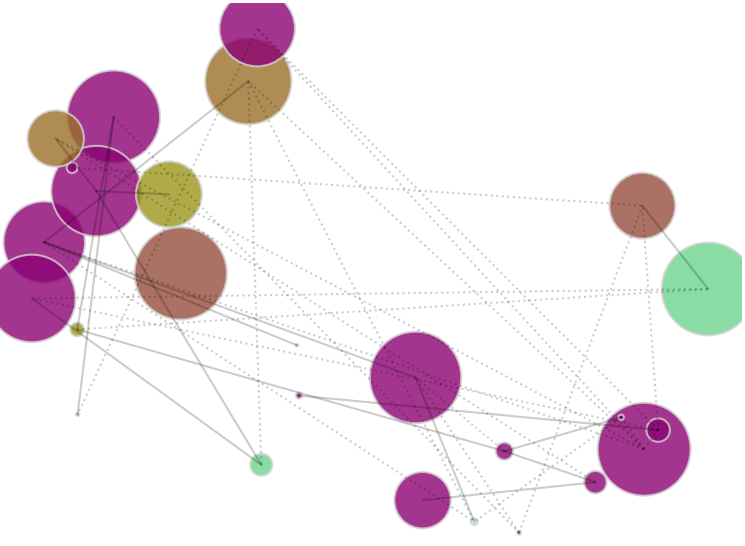}
    \label{fig:}
    }
\caption{\label{fig:arp}{\it \small Arpeggiator views between four notes indices. Colors represent note preferences.}}
\end{figure}

Arpeggiators select and play one note at a time from a set of user-pressed keys. Using real notes would require the network to understand the notion of interval, so the network generates an index, designated $\theta_t$ (Eq. \ref{eq:theta}), for a sorted list of notes.
The output is $n$-dimensional, with $n$ the number of keys pressed by the user: We initialize $\boldsymbol{W}_{out}$ with $m$ rows (maximum number of keys the user can press) but only activate $n$ rows at inference. The output is processed to select the index of the note to play as follows:
\begin{align}
    \boldsymbol{p}_t &= \text{softmax}(\beta \boldsymbol{y}_t) \\
    \theta_t &\sim \mathcal{CAT}(\theta_t | \boldsymbol{p}_t) \label{eq:theta}\\
    \boldsymbol{y}'_t &= \text{one\_hot}(\theta_t),
\end{align}
where $\beta \geq 0$ is dubbed the confidence parameter, $\mathcal{CAT}$ is the categorical distribution.
The user can tune the same hyperparameters as for the LFO to change dynamics and arpeggios. The confidence $\beta$ can be decreased to reduce determinism in the note selection process. The arpeggiator is illustrated in Fig. \ref{fig:arp}. 

\section{Conclusion}

We presented ReMi, a music production plugin suite based on random RNNs. ReMi improves upon standard arpeggiators and LFOs in two ways: signals are generated dynamically from the internal state of an RNN, not from pre-defined rules or hard-coded patterns; the user can adjust hyperparameters in real time to control output complexity, balancing determinism and randomness. We provided examples of oscillations and arpeggios generated with ReMi without any training. The LFO approach could be adapted to produce or transform audio signals. 
A related project \cite{steinmetz2020randomized} demonstrated interesting audio transformations using random convolutional neural networks (CNNs). Unlike CNNs, most ReMi hyperparameters are continuous, providing more intuitive control. 
Studies are being conducted to describe these effects formally.

One current limitation is the unsynchronized network dynamics with music software. While this is less critical for the arpeggiator, where the note selection rate can be externally controlled, synchronization with external signals would enhance both plugins, such as enabling the use of time signatures. One solution could be to introduce a rhythmic pulse as input, with user-controlled timing and intensity. Incorporating other user-defined input signals could provide more control over network dynamics, exceeding the limits of hyperparameter tuning alone. 

\begin{acknowledgments}
We thank all ReMi contributors. JUCE was used for the prototype plugins\footnote{\url{https://juce.com/}} and ReservoirPy \cite{Trouvain2020} used to develop Python prototypes.
\end{acknowledgments} 

\bibliography{ICMC2025_tmpl}

\begin{thebibliography}{1}
\providecommand{\url}[1]{#1}
\csname url@samestyle\endcsname
\providecommand{\newblock}{\relax}
\providecommand{\bibinfo}[2]{#2}
\providecommand{\BIBentrySTDinterwordspacing}{\spaceskip=0pt\relax}
\providecommand{\BIBentryALTinterwordstretchfactor}{4}
\providecommand{\BIBentryALTinterwordspacing}{\spaceskip=\fontdimen2\font plus
\BIBentryALTinterwordstretchfactor\fontdimen3\font minus \fontdimen4\font\relax}
\providecommand{\BIBforeignlanguage}[2]{{%
\expandafter\ifx\csname l@#1\endcsname\relax
\typeout{** WARNING: IEEEtran.bst: No hyphenation pattern has been}%
\typeout{** loaded for the language `#1'. Using the pattern for}%
\typeout{** the default language instead.}%
\else
\language=\csname l@#1\endcsname
\fi
#2}}
\providecommand{\BIBdecl}{\relax}
\BIBdecl

\bibitem{doshi2024generative}
A.~R. Doshi and O.~P. Hauser, ``Generative AI enhances individual creativity but reduces the collective diversity of novel content,'' \emph{Science Advances}, vol.~10, no.~28, p. eadn5290, 2024.

\bibitem{jaeger2004harnessing}
H.~Jaeger and H.~Haas, ``Harnessing nonlinearity: Predicting chaotic systems and saving energy in wireless communication,'' \emph{science}, vol. 304, no. 5667, pp. 78--80, 2004.

\bibitem{steinmetz2020randomized}
C.~J. Steinmetz and J.~D. Reiss, ``Randomized overdrive neural networks,'' \emph{arXiv preprint arXiv:2010.04237}, 2020.

\bibitem{Trouvain2020}
N.~Trouvain, L.~Pedrelli, T.~T. Dinh, and X.~Hinaut, ``{ReservoirPy}: An Efficient and User-Friendly Library to Design Echo State Networks,'' in \emph{Artificial Neural Networks and Machine Learning {\textendash} {ICANN} 2020}.\hskip 1em plus 0.5em minus 0.4em\relax Springer International Publishing, 2020, pp. 494--505.

\end{thebibliography}

\end{document}